\documentclass[conference]{IEEEtran}
\IEEEoverridecommandlockouts

\ifCLASSOPTIONcompsoc
\usepackage[caption=false, font=normalsize, labelfont=sf, textfont=sf]{subfig}
\else
\usepackage[caption=false, font=footnotesize]{subfig}
\fi

\usepackage{amsmath,amssymb,amsfonts}
\usepackage{algorithm}
\usepackage{algorithmic}
\renewcommand{\algorithmiccomment}[1]{\bgroup\hfill//~#1\egroup}
\usepackage{graphicx}
\usepackage{textcomp}
\usepackage{xcolor}

\usepackage{tikz}
\usetikzlibrary{shapes.geometric, arrows, dsp, chains}
\def\layersep{1.5cm}
\DeclareMathAlphabet{\mathpzc}{OT1}{pzc}{m}{it}

\usepackage[utf8]{inputenc}
\usepackage[english]{babel}

\usepackage[noadjust]{cite}

\usepackage{pgfplots}
\pgfplotsset{compat=newest}
\usepgfplotslibrary{groupplots}
\usepgfplotslibrary{dateplot}

\newcommand{\abs}[1]{\textcolor{red}{[abs: #1]}}

\usepackage[nolist,nohyperlinks]{acronym}

\definecolor{nn}{RGB}{35,139,69}
\definecolor{m2}{RGB}{254,178,76} 
\definecolor{m1}{RGB}{254,217,118} 
\definecolor{nn2}{RGB}{34,94,168} 
\definecolor{m4}{RGB}{252,197,192}
\definecolor{m4pq}{RGB}{228,26,28} 
\newcommand{\norm}[1]{\left\lVert#1\right\rVert}

\begin{document}
\bstctlcite{IEEEexample:BSTcontrol}

\title{Design and Implementation of a Neural Network Based Predistorter for Enhanced Mobile Broadband
}

\author{\IEEEauthorblockN{Chance Tarver\IEEEauthorrefmark{1},
		Alexios Balatsoukas-Stimming\IEEEauthorrefmark{2}\IEEEauthorrefmark{3}, and
		Joseph R. Cavallaro\IEEEauthorrefmark{1}
}%
	\IEEEauthorblockA{\IEEEauthorrefmark{1}Department of Electrical and Computer Engineering,
		Rice University, Houston, TX, USA}%
	\IEEEauthorblockA{\IEEEauthorrefmark{2}Department of Electrical Engineering,
		Ecole polytechnique f\'ed\'erale de Lausanne
		Lausanne, Switzerland}
	\IEEEauthorblockA{\IEEEauthorrefmark{3}Department of Electrical Engineering, 
		Eindhoven University of Technology, 
		Eindhoven, The Netherlands}%
}

\maketitle

\begin{abstract}
Digital predistortion is the process of correcting for nonlinearities in the analog RF front-end of a wireless transmitter. These nonlinearities contribute to adjacent channel leakage, degrade the \acl{EVM} of transmitted signals, and often force the transmitter to reduce its transmission power into a more linear but less power-efficient region of the device. Most predistortion techniques are based on polynomial models with an \acl{ILA} which have been shown to be overly sensitive to noise. In this work, we use \acl{NN} based predistortion with a novel \acl{NN} training method that avoids the \acl{ILA} and that shows significant improvements in both the \acl{ACLR} and \acl{EVM}. Moreover, we show that, by using a \acl{NN} based predistorter, we are able to achieve a 42\% reduction in latency and 9.6\% increase in throughput on an FPGA accelerator with 15\% fewer multiplications per sample when compared to a similarly performing memory-polynomial implementation.
\end{abstract}


\begin{acronym}
	\acro{DPD}{digital predistortion}
	\acro{PA}{power amplifier}
	\acro{RF}{radio frequency}
	\acro{TX}{transmit}
	\acro{MIMO}{multiple-input multiple-output}
	\acro{NN}{neural network}
	\acro{ILA}{indirect learning architecture}
	\acro{LS}{least squares}
	\acro{ML}{machine learning}
	\acro{FIR}{finite impule responce}
	\acro{PSD}{power spectral density}
	\acro{EVM}{error vector magnitude}
	\acro{RNN}{recurrent neural network}
	\acro{MSE}{mean squared error}
	\acro{ACLR}{adjacent channel leakage ratio}
	\acro{C-RAN}{cloud radio access network}
	\acro{GPU}{graphics processing units}
	\acro{SDR}{software defined radio}
	\acro{FCC}{Federal Communications Commission}
	\acro{ADC}{analog-to-digital converter}
	\acro{DAC}{digital-to-analog converter}
	\acro{IBFD}{in-band full-duplex}
	\acro{ReLU}{rectified linear unit}
	\acro{PAPR}{peak-to-average power ratio}
	\acro{PE}{processing element}
	\acro{LUT}{lookup table}
	\acro{FF}{flip-flop}
	\acro{DSP}{digital signal processor}
\end{acronym}

\section{Introduction}
Efficiently correcting nonlinearities  in \acp{PA} through \ac{DPD} is critical for enabling next-generation mobile broadband where there may be multiple \ac{RF} \ac{TX} chains arranged to form a massive \ac{MIMO} system~\cite{2014_MassiveMimo}, as well as new waveforms with bandwidths on the order of 100~MHz in the case of mmWave communications~\cite{2014_mmwave}. 
Traditional \acp{DPD} use variations of the Volterra series~\cite{2003_Volterra_DPD}, such as memory polynomials \cite{2010_JointMitigation, 2016_Katz}. These models consist of sums of various order polynomials and \ac{FIR} filters to model the nonlinearities and the memory effects in a \ac{PA}, respectively.

To learn the values of the parameters in a polynomial based model, an \ac{ILA} is typically used in conjunction with some variation of a \ac{LS} fit of the data to the model \cite{2016_Katz}. 
In an \ac{ILA}, a postinverse model of the predistorter is fitted based on the output of the \ac{PA}~\cite{Balatsoukas2015,Korpi2017}.
After learning the postinverter, the coefficients are copied to the predistorter.
Although this simplifies the learning of \ac{DPD} coefficients, it has been shown to converge to a biased solution due to noise in the PA output \cite{2007_ILA_IS_BAD, 2015_bias}.
Moreover, the \ac{LS} problem is often poorly conditioned \cite{2010_JointMitigation}.
%
In~\cite{2015_GPU_DPD}, a mobile \ac{GPU} was used to implement the polynomial \ac{DPD} with I/Q imbalance correction from~\cite{2010_JointMitigation}. 
This \ac{GPU} implementation used floating-point and was able to avoid the challenges associated with the dynamic range requirements for memory polynomials.
When implemented on an FPGA, a memory polynomial can be challenging due to the bit-widths that are necessary to perform the high-order exponentiation in fixed-point precision \cite{2011_Plume}.

The overall \ac{DPD} challenge has strong similarities to the problems encountered in \ac{IBFD} communications~\cite{Jain2011,Duarte2012,Bharadia2013}, where a transceiver simultaneously transmits and receives on the same frequency, increasing the spectral efficiency of the communication system. However, this requires (among other techniques) digitally removing the significant self-interference from the received signal which not only consists of the intended transmission but also the nonlinearities added by the imperfections in the transmit chain including the \ac{PA}.
In~\cite{2018_Alexios_FD_NN_Algo}, the authors used \acp{NN} to perform the self-interference cancellation and found that it could achieve similar performance to polynomial based self-interference cancellation. They later extended the work to create both FPGA and ASIC implementations of the \ac{NN}-based self-interference canceller and found that, due to the regular structure of the~\ac{NN} and the lower bit-width requirements, it can be implemented to have both a higher throughput and lower resource utilization~\cite{2018_Alexios_NN_FDD_Impl}.

Inspired by the full-duplex \ac{NN} work and the known problems of polynomial based predistortion with an \acp{ILA}, we recently proposed in~\cite{2019_Asilomar} to use \acp{NN} for the forward \ac{DPD} application. The NNs are a natural choice for such application as they are able to approximate any nonlinear function \cite{universal_approximation}, making them a reasonable candidate for predistortion. The idea of using various \acp{NN} for predistortion has been explored in many works \cite{2019_DNNDPD, 2012_NN_Ghannouchi}. However, the training method is unclear in~\cite{2019_DNNDPD}, and their implementations require over ten thousand parameters. In~\cite{2012_NN_Ghannouchi}, the training of the \ac{NN} is done using an \ac{ILA} which can subject the learned predistorter to the same problems seen with all \acp{ILA}.

\subsubsection*{Contribution} In our previous work~\cite{2019_Asilomar}, we avoided the standard \ac{ILA} 
and we improved the overall performance by using a novel training algorithm where we first modeled the \ac{PA} with a \ac{NN} and then backpropagated through it to train a \ac{DPD} \ac{NN}. We extend that work here to show that not only do we improve performance when compared to polynomial based \ac{DPD}, but we do so with reduced implementation complexity. Furthermore, to realize the gains of the \ac{NN} \ac{DPD}, we design a custom FPGA accelerator for the task and compare it to our own polynomial DPD accelerator.

\subsubsection*{Outline}
The rest of the paper is organized as follows. In Section II, we give an overview of our \ac{DPD} architecture and methods. In Section III, we compare performance/complexity tradeoffs for the \ac{DPD} \ac{NN} to polynomial based predistorters. In Section IV, we compare FPGA implementations for memory polynomial and \ac{NN} predistortion. Finally, in Section V we conclude the paper.

\section{Neural Network DPD Algorithm Overview}
 \begin{figure}[t]
	\centering	
	\begin{tikzpicture}
	\matrix (m1) [row sep=2.5mm, column sep=5mm]
	{
		\node[dspnodeopen,dsp/label=left] (start) {$x[n]$};  &
		\node[dspfilter]                   (nndpd) {\ NN DPD, $\hat{H}^{-1}$\ };  &
		\node[dspnodefull,dsp/label=above]                 (split) {$\hat{x}[n]$};        &
		\node[dspgain]                     (pa) {PA, $H$};         &
		\node[dspnodeopen, dsp/label=right](m06) {$y[n]$};       \\

		\node[coordinate]                  (row2_1) {};  &
		\node[coordinate]                  (row2_2) {};  &
		\node[coordinate]                  (row2_3) {};  &
		\node[coordinate]                  (row2_4) {};  &
		\node[dspmixer, dsp/label=right]               (row2_5) {$\frac{1}{G}$};  \\

		\node[coordinate]                  (m20) {};          &
		\node[coordinate]                  (m22) {};          &
		\node[coordinate]                  (m23) {};          &
		\node[dspfilter]                   (pann) {\ PA NN Model, $\hat{H}$\ };          &
		\node[coordinate]                  (m2X) {};    \\
	};
	
	
	\begin{scope}[start chain]
	\chainin (start);
	\chainin (nndpd) [join=by dspflow];
	\end{scope}
	
	\begin{scope}[start chain]
	\chainin (nndpd);
	\chainin (split) [join=by dspflow];
	\end{scope}
	
	\begin{scope}[start chain]
	\chainin (split);
	\chainin (pa) [join=by dspflow];
	\end{scope}
	
	\begin{scope}[start chain]
	\chainin (pa);
	\chainin (m06) [join=by dspflow];
	\end{scope}
	
	\draw [dashed] (m06) -- (row2_5);
	\draw [dashed] (row2_5) -- (m2X);
	\draw [->,dashed] (m2X) -- (pann);
	
	\draw [<-,dashed] (split) -- (m23);
	\draw [->,dashed] (m23) -- (pann);
	
	\node[align=right]  at (row2_3) {\textit{Training}};
	\end{tikzpicture}
	\caption{Architecture of the \ac{NN} \ac{DPD} system.
		The signal processing is done in the digital baseband and focuses on \ac{PA} effects. The \ac{DAC}, up/downconverters, and \ac{ADC} are not shown in this figure, though their impairments are also captured.} 
	\label{fig:NNsystem_arch}
\end{figure}
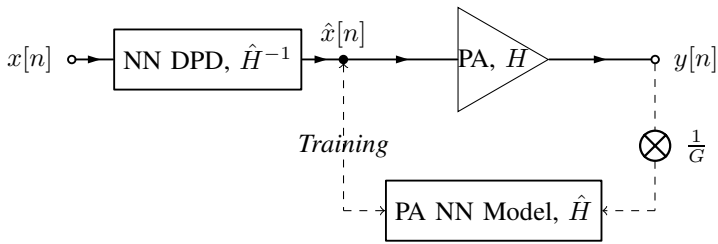
For the \ac{NN} \ac{DPD} system, we seek to place a \ac{NN} based predistorter inline with the \ac{PA} so that the cascade of the two is a linear system, as shown in Fig.~\ref{fig:NNsystem_arch}. 
However, to train a NN, it is necessary to have training data, and in this scenario the ideal \ac{NN} output is unknown; only the ideal \ac{PA} output is known. 
To overcome this problem, we train a \ac{PA} \ac{NN} model to emulate the \ac{PA}. 
We then backpropagate the \ac{MSE} through the \ac{PA} \ac{NN} model to update the parameters in the \ac{NN} \ac{DPD}~\cite{2019_Asilomar}.

\subsection{Neural Network Architecture}
We use a feed-forward \ac{NN} that is fully-connected with $K$ hidden layers, and $N$ neurons per hidden layer. The nonlinear activation applied in hidden layers is chosen to be a \ac{ReLU}, shown in~\eqref{eq:relu}, which can easily be implemented with a single multiplexer in hardware.
\begin{align}
\text{ReLU}(x) = \max(0,x)
\label{eq:relu}
\end{align}
The input and output data to the predistorter is complex-valued, while \acp{NN} typically operate on real-valued data. 
To accommodate this, we split the real and imaginary parts of each time-domain input sample, $x(n)$, on to separate neurons. 

\begin{figure}[t]
 	\centering
 	\begin{tikzpicture}[shorten >=1pt,->,draw=black!50, node distance=\layersep]
 	\tikzstyle{every pin edge}=[<-,shorten <=1pt]
 	\tikzstyle{neuron}=[circle,fill=black!25,minimum size=17pt,inner sep=0pt]
 	\tikzstyle{input neuron}=[neuron, fill=green!50];
 	\tikzstyle{output neuron}=[neuron, fill=red!50];
 	\tikzstyle{hidden neuron}=[neuron, fill=blue!50];
 	\tikzstyle{annot} = [text width=4em, text centered]
 	
 	\node[input neuron, pin=left:$\Re(x)$] (I-1) at (0, -0.5) {};
 	\node[input neuron, pin=left:$\Im(x)$] (I-2) at (0,-1.5) {};
 	
 	\foreach \name / \y in {1,...,2}
 	\path[yshift=0.5cm]
 	node[hidden neuron] (H-\name) at (\layersep,-\y cm) {};
 	
 	\node[align=left, rotate=90]  at (\layersep,-1 cm) {...};
 	\node[align=left]  at (\layersep+0.2cm,-1 cm) {$N$};

 	\node[output neuron,pin={[pin edge={->}]right:$\Re(y)$}, right of=H-1] (O-1) {};
 	\node[output neuron,pin={[pin edge={->}]right:$\Im(y)$}, right of=H-2] (O-2) {};
 	
 	\foreach \source in {1,...,2}
 	\foreach \dest in {1,...,2}
 	\path (I-\source) edge (H-\dest);
 	
 	\foreach \source in {1,...,2}
 	\path (H-\source) edge (O-1);
 	
 	\foreach \source in {1,...,2}
 	\path (H-\source) edge (O-2);

 	\path[bend left,->] (I-1) edge (O-1);
 	\path[bend right,->] (I-2) edge (O-2);
 	
 	\node[annot,above of=H-1, node distance=1cm] (hl) {Hidden layers};
 	\node[annot,left of=hl] {Input layer};
 	\node[annot,right of=hl] {Output layer};
 	\end{tikzpicture}
 	\caption{General structure of the \ac{DPD} and \ac{PA} neural networks. There are two input and output neurons for the real and imaginary parts of the signal, $N$ neurons per hidden layer, and $K$ hidden layers. The inputs are directly added to the output neurons so that the hidden layers concentrate on the nonlinear portion of the signal.}
 	\label{fig:NN_arch}
 \end{figure}
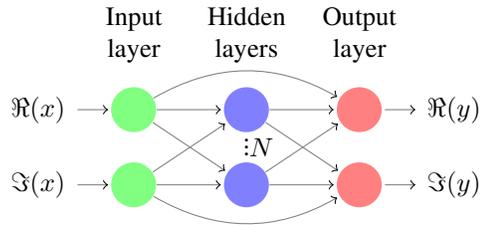

Although \ac{PA}-induced nonlinearities are present in the transmitted signal, the relationship between the input and output data is still mostly linear. 
Although in principle, a \ac{NN} can learn this relationship given training data, this turns out to be difficult in practice~\cite{2018_Alexios_FD_NN_Algo}. 
As such, we implement a linear bypass in our \ac{NN} that directly passes the inputs to the output neurons where they are added in with the output from the final hidden layer, as can be seen in Fig.~\ref{fig:NN_arch}. This way, the \ac{NN} entirely focuses on the nonlinear portion of the signal.

\subsection{Training}
This work primarily focuses on the implementation and running complexity of the \ac{DPD} application, which consists of inference on a pre-trained \ac{NN}. 
The training is assumed to be able to run offline and, once the model is learned, significant updates will not be necessary and occasional offline re-training to account for long-term variations would be sufficient.  

In~\cite{2019_Asilomar}, we first use input/output data of the \ac{PA} to train a \ac{NN} to model the \ac{PA} behavior. We then connect a second \ac{DPD} \ac{NN} to the \ac{PA} \ac{NN} model. We treat the combined \ac{DPD} \ac{NN} and \ac{PA} \ac{NN} as one large \ac{NN}. However, during the second training phase, we only update the weights corresponding to the \ac{DPD} \ac{NN}. We then connect the DPD NN to the real \ac{PA} and use it to predistort for the actual device. 

\begin{figure*}[t]
	\begin{align}
	\label{eq:mem_poly}
	\hat{x}(n) = &\sum_{\substack{p=1, \\p \text{ odd}}}^{P} \sum_{m=0}^{M} \alpha_{p, m} x(n-m) |  x(n-m) | ^{p-1} + \sum_{\substack{q=1, \\q \text{ odd}}}^Q \sum_{l=0}^{L} \beta_{q, l} x^*(n-l) |  x^*(n-l) | ^{q-1} + c
	\end{align}
	\hrule
\end{figure*}

\begin{figure}[t]
	\centering
\begin{tikzpicture}

\definecolor{color0}{rgb}{0.12156862745098,0.466666666666667,0.705882352941177}
\definecolor{color1}{rgb}{1,0.498039215686275,0.0549019607843137}

\begin{axis}[
width=6cm,
scale only axis,
legend cell align={left},
legend style={draw=white!80.0!black, font=\small},
tick align=outside,
tick pos=left,
title={NN MSE Training Loss},
x grid style={white!69.01960784313725!black},
xmajorgrids,
xlabel={Epoch},
xmin=-0.95, xmax=52,
xtick style={color=black},
y grid style={white!69.01960784313725!black},
ymajorgrids,
ylabel={MSE},
ymin=-0.00103111023985287, ymax=0.022034760006025,
ytick style={color=black}, 
scaled y ticks = false,
y tick label style={/pgf/number format/fixed}
]
\addplot [thick, color0]
table {%
0 0.0209863113584851
1 0.00868659276413183
2 0.00856596037600233
3 0.00847705272227541
4 0.00844331466807769
5 0.00843257975820511
6 0.00842832185375113
7 0.00842867227422331
8 0.00842712168136241
9 0.00842370320684635
10 0.00841998515191348
11 0.0084213544630315
12 0.00841925924229735
13 0.00841703796464836
14 0.00841282555064063
15 0.00841111319445788
16 0.00840676658761318
17 0.00840472792107435
18 0.00840056667210352
19 0.00840656997153125
};
\addplot [thick, color1]
table[x expr=\thisrowno{0}+20, y index=1] {%
0 0.00526144919076595
1 0.000108046817043831
2 3.62679249916481e-05
3 3.41087562847373e-05
4 3.36127235920356e-05
5 3.31138522086218e-05
6 3.27343582578073e-05
7 3.23281936186731e-05
8 3.1927089059979e-05
9 3.14678343744887e-05
10 3.10838720056683e-05
11 3.08800041979831e-05
12 3.03524742606431e-05
13 2.99466581017354e-05
14 2.95213201871469e-05
15 2.88490626652635e-05
16 2.83610668277233e-05
17 2.61337949091338e-05
18 1.75531967723105e-05
19 1.73384076870293e-05
};
\addplot [thick, color0]
table[x expr=\thisrowno{0}+40, y index=1] {%
	0 0.00839950884324356
	1 0.00839611467099636
	2 0.00839719791235196
	3 0.00839481377538423
	4 0.00839228366547754
};
\addlegendentry{PA NN Model}
\addplot [thick, color1]
table[x expr=\thisrowno{0}+45, y index=1] {%
	0 2.73791702273361e-05
	1 2.69200237014667e-05
	2 2.65309657863214e-05
	3 2.64952289731489e-05
	4 2.63441966303564e-05
};
\addlegendentry{DPD--PA NN}

\addplot +[mark=none, black, dashed] coordinates {(40, 0) (40, 1)};
\node[] at (axis cs: 10, 0.0125) {\small Iteration 1};
\node[] at (axis cs: 46,0.0125) {\small Iteration 2};
\end{axis}

\end{tikzpicture}
	\caption{Example of iterative NN-DPD training for two training iterations, where 20 and 5 epochs are used in the first and second iteration, respectively.}
	\label{fig:train}
\end{figure}
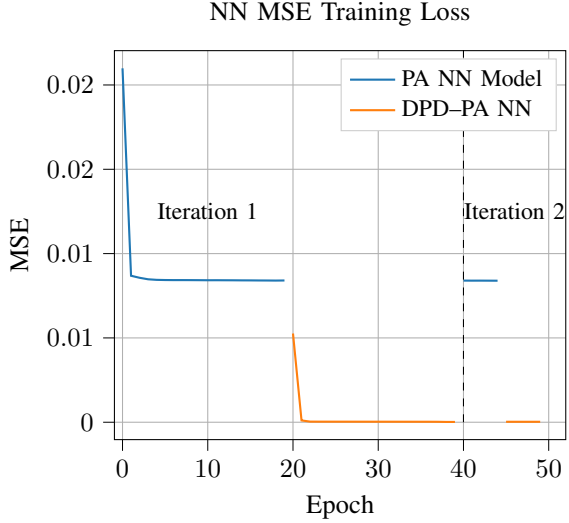

The process of predistorting can excite a different region of the \ac{PA} than when predistortion is not used. To account for this, it is not uncommon in other \ac{DPD} methods to have multiple training iterations. A similar idea is adopted in~\cite{2019_Asilomar} and in this work. Once training of the \ac{PA} and the \ac{DPD} is performed, we then retransmit through the actual \ac{PA} while using the \ac{DPD} \ac{NN}. 
Using the new batch of input/output data, we then can update the \ac{PA} NN model and in turn refine the \ac{DPD} \ac{NN}. 
An example of the iterative training procedure is shown in Fig.~\ref{fig:train}, where the \ac{MSE} training loss is shown for the \ac{PA} \ac{NN} model and the combined \ac{DPD}-\ac{PA} is shown for two training iterations.

\section{Complexity Comparison}
To evaluate the \ac{NN} based predistortion, we present the formulation of both a memory polynomial and the \ac{NN}. 
We then derive expressions for the number of multiplications as a function of the number of parameters in the models. 
In most implementations, multiplications are considered to be more expensive as they typically have higher latency and require more area and power. 
Additions typically have a minor impact on these metrics when compared to multiplications, so we omit them from this analysis.

\subsection{Memory Polynomial Predistortion}
An extension of a memory polynomial from \cite{2010_JointMitigation} is shown in~\eqref{eq:mem_poly}. 
This form of memory polynomial predistorts the complex baseband \ac{PA} input $x(n)$ to be $\hat{x}(n)$ by computing nonlinearities of the form $x(n)|x(n)|^p$ and convolving them with an \ac{FIR} filter for both $x(n)$ and its conjugate, $x^*(n)$. 
This conjugate processing gives the model the expressive power to combat \ac{PA} nonlinearities and any IQ imbalance in the system. 
$P$ and $M$ are the highest nonlinearity order and memory depth in the main branch, while $Q$ and $L$ are the highest order and memory in the conjugate branch. 
The complex-valued coefficients $\alpha_{p,m}$ and $\beta_{q,l}$ represent the \ac{DPD} coefficients that need to be learned for nonlinearity orders $p$ and $q$ and memory tap $m$ and $l$. 
Finally, the DC term $c$ accounts for any local oscillator leakage in the system.

The total number of complex-valued parameters in~\eqref{eq:mem_poly} is given as 
\begin{align}
n_{\text{PAR, poly}} = M\left(\frac{P+1}{2} \right) +  L\left(\frac{Q+1}{2} \right) + 1.
\end{align}
Assuming three real multiplications per complex multiplication, we get the following number of multiplications in the system
\begin{align}
n_{\text{MUL, poly}} &= 3 n_{\text{PAR, poly}} 
+ \sum_{\substack{p=3, \\p \text{ odd}}}^{P} \frac{1}{2}\left(p+5\right)
+ \sum_{\substack{q=3, \\q \text{ odd}}}^{Q} \frac{1}{2}\left(q+5\right)
\end{align}
Here, each complex coefficient accounts for three multiplication. The expression, $x(n)|x(n)|^{p-1}$ is computed once for each $n$ over a given $p$ and delayed in the design to generate the appropriate value for each $m$. We note that $|x(n)|^{p-1}$ can always be simplified to $(\Re{(x(n))^2 + \Im{(x(n)}}^2)^{\frac{p-1}{2}}$ since $p$ is odd. 
This accounts for $ (\frac{p-1}{2} + 1)$ multiplications before being multiplied by the complex-valued $x(n)$ which adds 2 more multiplications. The same is true for the conjugate processing.

\subsection{Neural Network Predistortion}
The output of a densely connected NN is given by
\begin{align}
\label{eq:nn_h1}
{\mathbf{h}_1}(n) = f\left(\mathbf{W}_1
\begin{bmatrix}
\Re(x(n))\\
\Im(x(n))
\end{bmatrix}
+ \mathbf{b}_1
\right),
\end{align}
\begin{align}
\label{eq:nn_hi}
{\mathbf{h}_i}(n) = f\left(\mathbf{W}_i
\mathbf{h}_{i-1}(n) + \mathbf{b}_i\right), \quad i = 2,\hdots,K,
\end{align}
\begin{align}
\label{eq:nn_out}
\mathbf{z}(n) = \mathbf{W}_{K+1} \mathbf{h}_{K}(n) + \mathbf{b}_{K+1} + \mathbf{W}_{\text{linear}}
\begin{bmatrix}
\Re(x(n))\\
\Im(x(n))
\end{bmatrix},
\end{align}
\begin{align}
\hat{x}(n) = z_1(n) + 1j \cdot z_2(n),
\end{align}
where $f$ is a nonlinear activation function (such as the \ac{ReLU} from~\eqref{eq:relu}), $\mathbf{W}_i$ and $\mathbf{b}_i$ are weight matrices and bias vectors corresponding to the $i$th layer in the \ac{NN}, and $j$ is the imaginary unit. The final output of the network after hidden layer $K$ is given by~\eqref{eq:nn_out} where the first element represents the real part of the signal, and the second element represents the imaginary part.
In~\eqref{eq:nn_out}, $W_{\text{linear}}$ is a 2$\times$2 matrix of the weights corresponding to the linear bypass. In practice, we fix it to be the identity matrix, $\mathbf{I}_2$, to reduce complexity though these weights could also be learned in systems with significant IQ imbalance.

Assuming $N$ neurons per hidden layer and $K$ hidden layers, the number of multiplications is given by
\begin{align}
n_{\text{MUL, NN}} = 4N + (K-1)N^2.
\end{align}

\subsection{Results}
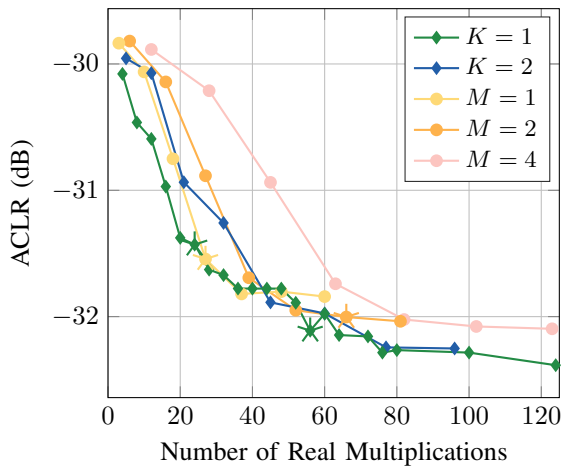
\begin{figure}
	\centering
%
%
\definecolor{mycolor1}{rgb}{0.00000,0.44700,0.74100}%
\begin{tikzpicture}

\begin{axis}[%
reverse legend,
width=6cm,
scale only axis,
xmin=0,
xmax=125,
xmajorgrids,
ymajorgrids,
axis background/.style={fill=white},
xlabel={Number of Real Multiplications},
ylabel={ACLR (dB)},
legend style={legend cell align=left, align=left, draw=white!15!black, font=\small},
every axis plot/.append style={thick}
]
\addplot [color=m4, mark=*, mark options={solid, m4}]
table[row sep=crcr]{%
	12	-29.8853\\ 	
	28	-30.2118\\	
	45	-30.9373\\	
	63	-31.7399\\	
	82	-32.0228\\	
	102	-32.0779\\	
	123	-32.0963\\	
};
\addlegendentry{$M=4$}

\addplot [color=m2, mark=*, mark options={solid,m2}]
table[row sep=crcr]{%
	6	-29.818862527181\\
	16	-30.1421895606294\\
	27	-30.8850600348483\\
	39	-31.692266157724\\
	52	-31.9493882490671\\
	66	-32.0030872098609\\
	81	-32.0369669401838\\
};
\addlegendentry{$M=2$}

\addplot [color=m1, mark=*, mark options={solid,m1}]
  table[row sep=crcr]{%
3	-29.8356\\
10	-30.0628\\
18	-30.7505\\
27	-31.5441\\
37	-31.8203\\
48	-31.8019\\
60	-31.8413\\
};
\addlegendentry{$M=1$}

\addplot [color=nn2, mark=diamond*, mark options={solid,nn2}]
table[row sep=crcr]{%
	5 	-29.95584291149705\\	
	12	-30.072466004072524\\	
	21	-30.936133857555134\\ 
	32	-31.25849680582959\\	
	45	-31.887107955929835\\
	60	 -31.974556494548914\\
	77	 -32.243370057951594\\
	96	-32.252043179348306\\
};
\addlegendentry{$K=2$}

\addplot [color=nn, mark=diamond*, mark options={solid,nn}]
table[row sep=crcr]{%
	4 	-30.078717636746653\\
	8	-30.463175118195153\\	
	12	-30.593144365075805\\ 	
	16	-30.970344493134434\\ 	
	20	-31.376524622879497\\  	
	24	-31.42988155703645\\	
	28	-31.62784377719193\\	
	32	-31.671292599866717\\	
	36	-31.77752710075101\\	
	40	-31.778105475618226\\	
	44	-31.778105475618226\\	
	48	-31.778105475618226\\	
	52	-31.890500108655054\\	
	56	-32.11119371251352\\	
	60	-31.977260126552103\\   
	64	-32.14583884446011		
	68	-32.17262237702159\\
	72	-32.156610658061965\\
	76	-32.28473318192649\\
	80	-32.26490257813185\\
	100	-32.28473318192649\\ 
	124	-32.384432317036676\\
	240	-32.45072645283004\\
};
\addlegendentry{$K=1$}

\addplot [mark=star, mark size=5pt, mark options={solid,nn}, forget plot]
coordinates {(24,	-31.42988155703645)};
\addplot [mark=star, mark size=5pt, mark options={solid,m1}, forget plot]
coordinates {(27, -31.5441)};

\addplot [mark=star, mark size=5pt, mark options={solid,nn}, forget plot]
coordinates {(56, -32.11119371251352)};
\addplot [mark=star, mark size=5pt, mark options={solid,m2}, forget plot]
coordinates {(66, -32.0030872098609)};

\end{axis}
\end{tikzpicture}%
	\caption{\ac{ACLR} vs. number of multiplications for \ac{NN} \ac{DPD} (shown with diamonds) with up to $K=2$ hidden layers and memory polynomial (shown with circles) with up to $M=4$ memory taps. This represents the out-of-band performance of the predistorter. The stars represent design points that we implement in FPGA in the next section.}
	\label{fig:aclr}
\end{figure}

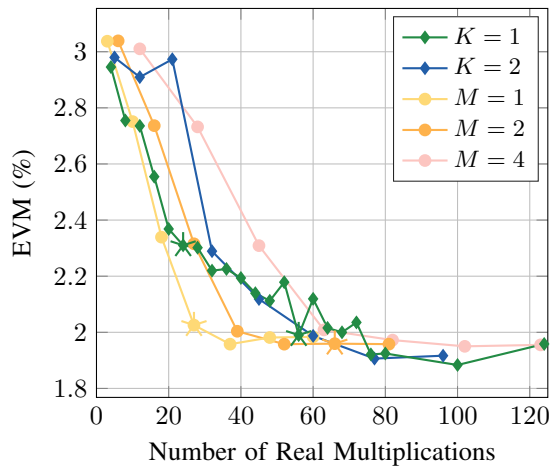
\begin{figure}
	\centering
%
%
\definecolor{mycolor1}{rgb}{0.00000,0.44700,0.74100}%
\begin{tikzpicture}

\begin{axis}[%
reverse legend,
width=6cm,
scale only axis,
xmin=0,
xmax=125,
xmajorgrids,
ymajorgrids,
axis background/.style={fill=white},
xlabel={Number of Real Multiplications},
ylabel={EVM (\%)},
legend style={legend cell align=left, align=left, draw=white!15!black, font=\small},
every axis plot/.append style={thick}
]

\addplot [color=m4, mark=*, mark options={solid,m4}]
table[row sep=crcr]{%
	12	3.0103\\
	28	2.7322\\
	45	2.3089\\
	63	2.0085\\
	82	1.9722\\
	102	1.95\\
	123	1.9552\\
};
\addlegendentry{$M=4$}

\addplot [color=m2, mark=*, mark options={solid,m2}]
table[row sep=crcr]{%
	6	3.0388192857864\\
	16	2.73656091785756\\
	27	2.3154671075041\\
	39	2.00375004718572\\
	52	1.95769885368691\\
	66	1.95947582666292\\
	81	1.95781082858679\\
};
\addlegendentry{$M=2$}

\addplot [color=m1, mark=*, mark options={solid, m1}]
table[row sep=crcr]{%
	3	3.0375\\
	10	2.751\\
	18	2.3392\\
	27	2.0256\\
	37	1.9577\\
	48	1.9814\\
	60	1.9837\\
};
\addlegendentry{$M=1$}

\addplot [color=nn2, mark=diamond*, mark options={solid,nn2}]
table[row sep=crcr]{%
	5 	2.9797\\	
	12	2.9095\\	
	21	2.9732\\	
	32	2.2891\\	
	45	2.1186\\
	60	 1.9872\\
	77	1.9063\\
	96 	1.9166\\
};
\addlegendentry{$K=2$}

\addplot [color=nn, mark=diamond*, mark options={solid, nn}]
table[row sep=crcr]{%
	4	2.9454\\ 	
	8	2.7549\\ 	
	12	2.7357\\ 	
	16	2.5547\\ 	
	20	2.3683\\ 	 	
	24	2.3100\\	
	28	2.3016\\	
	32	2.2198\\	
	36	2.2257\\	
	40	2.1936\\	
	44	2.1394\\	
	48	2.1122\\	
	52	2.1786\\	
	56	1.99\\
	60	2.1192\\
	64	2.0156\\
	68	 1.9999\\
	72	2.0352\\
	76	1.9212\\
	80	1.9245\\
	100	1.8835\\
	124	1.9584\\
	240	 1.8988\\
};
\addlegendentry{$K=1$}

\addplot [mark=star, mark size=5pt, mark options={solid,nn}, forget plot]
coordinates {(24,	2.310)};
\addplot [mark=star, mark size=5pt, mark options={solid,m1}, forget plot]
coordinates {(27, 	2.0256)};

\addplot [mark=star, mark size=5pt, mark options={solid,nn}, forget plot]
coordinates {(56, 1.99)};
\addplot [mark=star, mark size=5pt, mark options={solid,m2}, forget plot]
coordinates {(66, 1.95947582666292)};

\end{axis}
	\end{tikzpicture}%
	\caption{\ac{EVM} vs. number of real multiplications for \ac{NN} \ac{DPD} (shown with diamonds) with up to $K=2$ hidden layers and memory polynomial (shown with circles) with up to $M=4$ memory taps. This represents the in-band performance of the predistorter.  The stars represent design points that we implement in FPGA in the next section}
	\label{fig:evm}
\end{figure}

\begin{figure}
	\centering
	\input{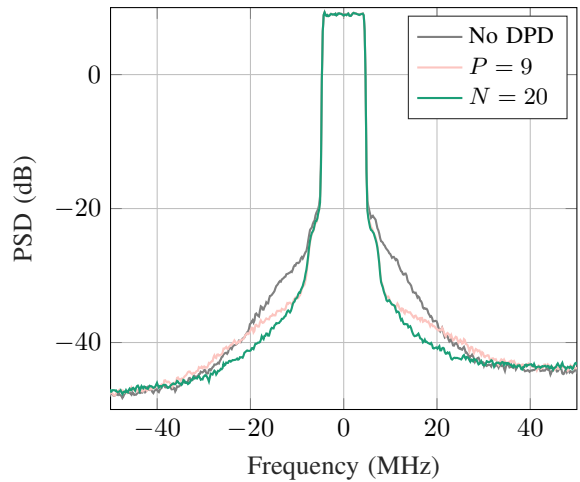}
	\caption{Example spectrum for the $M=4$ polynomial and $K=1$ \ac{NN}. Each of these use around 80 multiplications per time-domain input sample to the \ac{DPD}.}
	\label{fig:psd}
\end{figure}

The performance results for each predistorter as a function of the number of required multiplications are shown in Figs.~\ref{fig:aclr}--\ref{fig:psd}. These results were obtained using the RFWebLab platform \cite{rfweblab}. 
RFWebLab is a web-connected PA at Chalmers University.
This system uses a Cree CGH40006-TB GaN PA with a peak output power of 6 W. 
The precision is 14 bits for the feedback on the \ac{ADC} and 16 bits for the \ac{DAC}. 
Using their \textsc{Matlab} API, we test the \ac{NN} predistorter using a 10~MHz OFDM signal.
This signal has random data on 600 subcarriers spaced apart by 15~kHz and is similar to LTE signals commonly used in cellular deployments. It provides an interesting test scenario in that it has a sufficiently high \ac{PAPR} to make predistortion challenging. 
We train on 10 symbols then validate on 10 different symbols. 
The Adam optimizer is used with an \ac{MSE} loss function. 
\ac{ReLU} activation functions are used in the hidden layer neurons. 

Specifically, we tested the following \acp{DPD}: 
1) a \ac{NN} DPD with $K=1$ with $N=\{1,..., 20, 25, 31\}$ (dark green). 
2) a \ac{NN} DPD with $K=2$ with $N=\{1,...,8\}$ (light green).
3) a polynomial \ac{DPD} without memory and with $P=1$ to $P=13$ (dark blue),
2) a polynomial \ac{DPD} with $M=2$ memory taps and with $P=1$ to $P=13$ (light blue), 
3) a polynomial \ac{DPD} with $M=4$ memory taps and with $P=1$ to $P=13$ (pink), 
All \acp{DPD} were evaluated in terms of the \ac{ACLR}, the \ac{EVM}, and the spectra of the post-PA pre-distorted signals.
A predistorter with $M=4$ and $Q=P$ was also evaluated. However, the system did not have significant IQ imbalance, so the addition of the conjugate processing to the memory polynomial only had the effect of significantly increasing complexity.


\subsubsection{Out-of-band performance}
To measure the out-of-band performance, which is often the metric of most interest given by \ac{FCC} regulations and 3GPP standards, we compute the \ac{ACLR} shown below as
\begin{align}
	\text{ACLR} = 10 \log_{10}\frac{P_{\text{adjacent}}}{P_{\text{channel}}},
	\label{eq:aclr}
\end{align}
where $P_{\text{channel}}$ is the signal power in the main channel, and $P_{\text{adjacent}}$ is the signal power in the remainder of the band. 

In Fig.~\ref{fig:aclr}, we observe that the \ac{NN} \ac{DPD} offers similar performance to the memoryless polynomial \ac{DPD} for low numbers of multiplications and it is able to significantly outperform all polynomial \acp{DPD} as the number of multiplications increases. 

\subsubsection{In-band performance}
Although the primary goal of predistortion is to reduce spectral regrowth around the main carrier, predistortion also reduces the \ac{EVM} of the main signal. Reducing \ac{EVM} can improve reception quality and is hence a desirable result. 
The \ac{EVM} is computed as 
\begin{align}
\text{EVM} = \frac{\norm{\hat{\mathbf{s}} - \mathbf{s}}}{\norm{\mathbf{s}} } \times 100\%,
\label{eq:evm}
\end{align}
where $\mathbf{s}$ is the vector of all original symbols mapped onto complex constellations on OFDM subcarriers in the frequency domain, $\hat{\mathbf{s}}$ is the corresponding received vector after passing through the \ac{PA}, and $\norm{\cdot}$ represents the $\ell^2$ norm.

In Fig.~\ref{fig:evm}, we see the \ac{EVM} versus the number of multiplications for each of the predistorters. 
As the number of multiplications increases, the \ac{EVM} decreases, as expected. 
The memoryless polynomial \ac{DPD} is able to achieve a low \ac{EVM} for the smallest number of multiplications. 
However, the complexity is only slightly higher for the \ac{NN} based \ac{DPD}, which is able to achieve an overall better performance than all other examined polynomial \acp{DPD}.

\subsubsection{Spectrum Comparison}
The spectrum for both the memory polynomial and the \ac{NN} \acp{DPD} are shown in Fig. \ref{fig:psd}. Here, both predistorters have the same running complexity of 80 multiplications per time-domain input sample. However, the \ac{NN} is able to provide an additional 2.8 dB  of suppression at $\pm20$ MHz.

\section{FPGA Architecture Overview}
In this section, we compare a \ac{NN} \ac{DPD} accelerator with a memory polynomial based implementation. We implement both designs in Xilinx System Generator and target for the Zynq UltraScale+ RFSoC ZCU1285 evaluation board. For the sake of this architecture comparison, we implement each to be fully parallelized and pipelined as to compare the highest throughput implementations of each. Based on the previous analysis, we implement both with 16-bit fixed point precision throughout. 

We synthesize FPGA designs targeting two separate \acp{ACLR}. First, we target an \ac{ACLR} of approximately -31.4 dB. 
This target is achieved with a \ac{NN} with $N=6$ neurons and $K=1$ hidden layer and a 7th order memoryless polynomial. 
Second, we target a more aggressive \ac{ACLR} below -32 dB. This is done with a \ac{NN} with $N=14$ neurons and $K=1$ hidden layer. A memory polynomial with $M=2$ and $P=11$ is also used to achieve this.

\subsection{Neural Network Accelerator}
We implement the \ac{NN}-DPD on FPGA with the goal of realizing high throughput via maximum parallelization and pipeling. 
The top-level overview of the design is shown in Fig. \ref{fig:nn_fpga}.
Here, each wire corresponds to a 16-bit bus. 
The real and imaginary parts of the \ac{PA} input signal stream in each clock cycle. 
Weights are stored in a RAM which can be written to from outside the FPGA design.
After the RAM is loaded, the weights and biases are written to individual registers in the neuron \acp{PE} which cache them for fast access during inference. A chain of pipeline registers pass the inputs to the output to be added to the output of the final layer.

After the weights are loaded into RAM, the RAM controller loads each of the weights into a weights cache in each \ac{PE}.
To do this, a counter increments through each address in the RAM. The current address and the value at that address are broadcast to all neurons. Each address corresponds with a specific weight or bias. Whenever the weights cache in a neuron reads addresses corresponding to the weights and biases for its neuron, it saves the data into a register dedicated to that parameter. These registers output to the corresponding multiplier or adder.

An example neuron \ac{PE} is shown in Fig. \ref{fig:pe_fpga}.
Each \ac{PE} is implemented with a sufficient number of multipliers for performing the multiplication of the weights by the inputs in parallel. 
The results from each multiplier are added together, along with the bias and passed to the \ac{ReLU} activation function, which is implemented with a single multiplexer. 

\begin{figure}
	\centering
	\begin{tikzpicture}[
roundnode/.style={circle, draw=green!60, fill=green!5, very thick, minimum size=7mm},
squarednode/.style={rectangle, align=center, draw=black, very thick, minimum size=6mm, text width=6mm},
node distance=0.25cm and 1cm,
]
\node(real_input)						{\small $\Re(x[n])$};
\node(imag_input)[below=0.31cm of real_input]	{\small $\Im(x[n])$};

\node[squarednode](pe1)[right=of real_input]      {\small PE};
\draw[->] (real_input.east) -- (pe1.west);
\draw[->] (imag_input.east) -- (pe1.west);

\node[squarednode](pe2)[below=of pe1]      {\small PE};
\draw[->] (real_input.east) -- (pe2.west);
\draw[->] (imag_input.east) -- (pe2.west);

\node(dots_h1)[below=of pe2, rotate=90]{...};

\node[squarednode](peN)[below=0.5cm of pe2] {\small PE};
\draw[->] (real_input.east) -- (peN.west);
\draw[->] (imag_input.east) -- (peN.west);

\node[squarednode](linear)at(3.2,-3) [text width = 25mm] {\small Linear Bypass Pipeline Registers};
\draw[->] (real_input.east) -- ++(0.5cm,0) |- ([yshift=0.1cm]linear.west);
\draw[->] (imag_input.east) -- ++(0.4cm,0) |- (linear.west);

\node[squarednode, draw=black, very thick](out1)[right=of pe1]{\small PE};
\draw[->] (pe1.east) -- (out1.west);
\draw[->] (pe2.east) -- (out1.west);
\draw[->] (peN.east) -- (out1.west);

\node[squarednode, draw=black, very thick](out2)[right=of pe2]{\small PE};
\draw[->] (pe1.east) -- (out2.west);
\draw[->] (pe2.east) -- (out2.west);
\draw[->] (peN.east) -- (out2.west);

\node[squarednode, draw=black, very thick](add1)[right=0.1cm of out1]{\small Add};
\draw[->] (out1.east) -- (add1.west);
\draw[->] ([yshift=0.1cm]linear.east) -| (add1.south);

\node[squarednode, draw=black, very thick](add2)[right=0.65of out2]{\small Add};
\draw[->] (out2.east) -- (add2.west);
\draw[->] (linear.east) -| (add2.south);

\node(output)[right=0.5cm of add1]	{$\Re(\hat{x}[n])$};
\draw[->] (add1.east) -- (output.west);

\node(output)[right=0.1cm of add2]	{$\Im(\hat{x}[n])$};
\draw[->] (add2.east) -- (output.west);

\node[rectangle, draw=black, text width=22mm, align=center,very thick](ram)[above=of pe1]  {\small Weights and Biases RAM};
\draw[->, dashed] (ram.south) -- (pe1.north);
\draw[->, dashed] (pe1.south) -- (pe2.north);
\draw[->, dashed] (pe2.south) -- (peN.north);
\draw[->, dashed] (ram.east) -| (out1.north);
\draw[->, dashed] (out1.south) -- (out2.north);
\end{tikzpicture}
	\caption{General structure of the \ac{NN} FPGA implementation.}
	\label{fig:nn_fpga}
\end{figure}
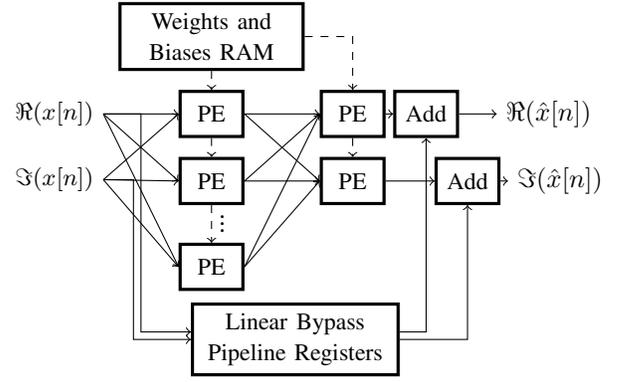
\begin{figure}
	\centering
	\begin{tikzpicture}[
roundnode/.style={circle, draw=green!60, fill=green!5, very thick, minimum size=7mm},
squarednode/.style={rectangle, align=center, draw=black, very thick, minimum size=6mm, text width=12mm},
node distance=0.25cm and 1cm,
]
\node(real_input)at (0,0)					{\small $\Re(x[n])$};
\node(from_ram)[below=0.5cm of real_input]	{\small From RAM};
\node(imag_input)[below=0.5cm of from_ram]	{\small $\Im(x[n])$};

\node[rectangle, draw=black, very thick, text width = 1.3cm, align=center](logic)[right=0.77cm of from_ram]{\small Weights Cache};
\draw[->] (from_ram.east) -- (logic.west);

\node[rectangle, draw=black, very thick, text width = 1.3cm, align=center](mult_1)[right=1cm of real_input]{\small Multiply};
\node[rectangle, draw=black, very thick, text width = 1.3cm, align=center](mult_2)[right=1cm of imag_input]{\small Multiply};

\draw[->] (real_input.east) -- (mult_1.west);
\draw[->] (imag_input.east) -- (mult_2.west);

\node[rectangle, draw=black, very thick](add)[right=0.6cm of mult_1]{\small Add};
\draw[->] (mult_1.east) -- (add.west);

\node[rectangle, draw=black, very thick](add_bias)[right=0.6cm of logic]{\small Add};
\draw[->] (mult_2.east) -| (add_bias.south);
\draw[->] (logic.east) -- (add_bias.west);
\draw[->] (add_bias.north) -| (add.south);

\node[rectangle, draw=black, very thick, text width=0.8cm](relu)[right=0.4cm of add]{\small ReLU Mux};
\draw[->] (add.east) -- (relu.west);

\node(neuron_out)[right=0.4cm of relu]				{\small $h_{1,i}(n)$};
\draw[->] (relu.east) -- (neuron_out.west);

\draw[->] (logic.north) -- (mult_1.south);
\draw[->] (logic.south) -- (mult_2.north);
\end{tikzpicture}
	\caption{Example structure of a \ac{PE} for the $i$th neuron in hidden layer 1.}
	\label{fig:pe_fpga}
\end{figure}
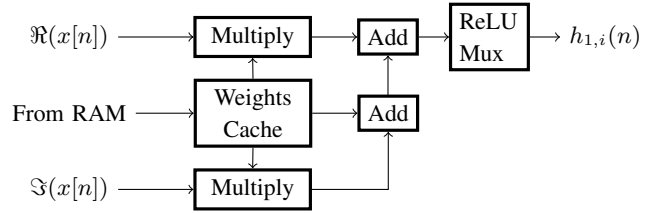

\subsection{Polynomial Accelerator}
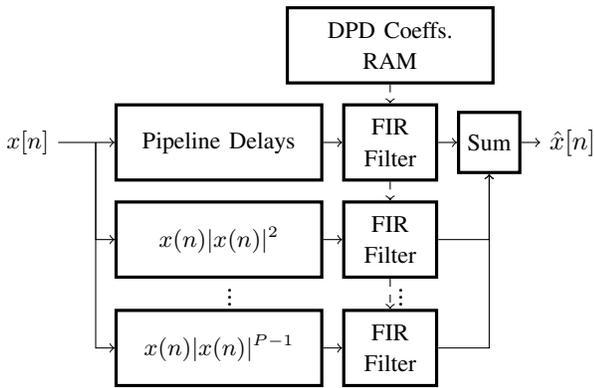
\begin{figure}
	\centering
	\begin{tikzpicture}[
roundnode/.style={circle, draw=green!60, fill=green!5, very thick, minimum size=7mm},
squarednode/.style={rectangle, align=center, draw=black, very thick, minimum size=10mm, text width=25mm},
node distance=0.25cm and 0.25cm,
]
\node(preinput)	{\small $x[n]$};
\node(input)[right=of preinput]{};
\draw[-] (preinput.east) -- (input.east);

\node[squarednode](p1)[right=of input]  {\small Pipeline Delays};
\node[squarednode](p1fir)[right=of p1, text width=10mm]      {\small FIR Filter};
\draw[->] (input.east) -- (p1.west);
\draw[->] (p1.east) -- (p1fir.west);

\node[squarednode](p3) [below=of p1] {\small $x(n)|x(n)|^2$};
\node[squarednode](p3fir)[below=of p1fir,  text width=10mm]      {\small FIR Filter};
\draw[->] (p3.east) -- (p3fir.west);
\draw[->] (input.east) |- (p3.west);

\node(dots)[below=of p3, rotate=90]{...};
\node(dotsfir)[below=of p3fir, rotate=90]{...};

\node[squarednode](pP) [below=0.4cm of p3] {\small $x(n)|x(n)|^{P-1}$};
\node[squarednode](pPfir)[below=0.4cm of p3fir,  text width=10mm] {\small FIR Filter};
\draw[->] (pP.east) -- (pPfir.west);
\draw[->] (input.east) |- (pP.west);

\node[rectangle, draw=black, very thick, minimum size=8mm,](sum)[right=of p1fir]{\small Sum};
\draw[->] (p1fir.east) -- (sum.west);
\draw[->] (p3fir.east) -| (sum.south);
\draw[->] (pPfir.east) -| (sum.south);

\node(output)[right=of sum]	{$\hat{x}[n]$};
\draw[->] (sum.east) -- (output.west);

\node[squarednode](ram)[above=of p1fir]  {\small DPD Coeffs. RAM};
\draw[->, dashed] (ram.south) -- (p1fir.north);
\draw[->, dashed] (p1fir.south) -- (p3fir.north);
\draw[->, dashed] (p3fir.south) -- (pPfir.north);
\end{tikzpicture}
	\caption{General structure of the high-throughput, low-latency, memory polynomial FPGA implementation. }
	\label{fig:poly_fpga}
\end{figure}

The memory polynomial is also implemented using 16 bits throughout the design. 
We target the design for maximum throughput by fully parallelizing and pipelining it so that a new time-domain input sample can streamed in each clock cycle.
The main overall structure of the design is shown in Fig. \ref{fig:poly_fpga}.
Each polynomial ``branch" of the memory polynomial corresponding to  nonlinear order $p$ computes $x(n)|x(n)|^{p-1}$ and there is a branch for each $p$ in the design. 
This computation from each branch is passed to an FIR filter with complex taps. 
Three multiplications are used for each complex multiplication in each filter.
A RAM is implemented to interface with some outside controller for receiving updated weights. 
Once the coefficients $\alpha$ and $\beta$ are loaded into the design, they can be moved from the RAM to registers near each multiply similarly to the cache implemented in the \ac{NN} design.

\subsection{Results}
The Xilinx Vivado post-place-and-route  utilization results are shown in Table \ref{tab:nn}.  
Overall, the \ac{NN}-based design offers numerous advantages over the memory polynomial.
Specifically, for the target of an \ac{ACLR} less than -32 dB, the \ac{NN} requires 48\% of the \acp{LUT},   42\% of the \acp{FF}, and 15\% reduction in the number of \acp{DSP}. In terms of timing, there is a 9.6\% increase in throughput with a 46\% decrease in latency. These reductions in utilization occur while also seeing improved \ac{ACLR}.

\newcommand{\specialcell}[2][c]{%
	\begin{tabular}[#1]{@{}c@{}}#2\end{tabular}}
\begin{table}
	\caption{Comparison of Performance and FPGA Utilization}
	\label{tab:nn}
	\centering
	\begin{tabular}{p{2.5cm}| p{0.9cm} | p{0.9cm} |p{0.9cm} | p{0.9cm}}
		 &  \multicolumn{2}{c|}{ACLR: -31.4 dB} & \multicolumn{2}{c}{ACLR: -32} \\
		\centering Metric  				&\specialcell{$N=6$\\$K=1$}	& \specialcell{$P=7$\\$M=1$}& \specialcell{$N=14$\\$K=1$} & \specialcell{$P=11$\\$M=2$}	\\ \hline
		Num. of Params.			& 32		& 8		& 72	& 24	\\
		\hline
		LUT						& 379		& 539	& 688	& 1424\\
		LUTRAM					& 16		& 120	& 16	& 224\\
		FF						& 538		& 991	& 1170	& 2730\\
		DSP						& 24		& 27	& 56	& 66\\
		\hline
		Worst Neg. Slack (ns)	& 8.72 		& 8.68	& 8.49	& 8.34 \\
		Max. Freq. (MHz)		& 783		& 756	& 661	& 603	\\ 
		Max. T/P (MS/s)			& 783		& 756	& 661	& 603	\\ 
		Latency (CC)			& 12 		& 21	& 14	& 26
	\end{tabular}
\end{table}

\section{Conclusions}
In this paper, we explored the complexity/performance tradeoffs for a novel, \ac{NN} based \ac{DPD} and found that the \ac{NN} could outperform memory polynomials and offered overall unrivaled \ac{ACLR} and \ac{EVM} performance. Furthermore, we implemented each on an FPGA and found that the regular matrix multiply structure in the NN based predistorter led to a lower latency design with less hardware utilization when compared to a similarly performing polynomial-based DPD.

This work opens up many avenues for future work. This work can be extended to also compare performance/complexity tradeoffs for more devices with a wider variety of signals, including different bandwidths and multiple component carriers. It is also possible to include memory cells such as \acp{RNN} in the \ac{NN} to account for memory effects. The \ac{NN} is naturally well suited for a GPU implementation which would be interesting in \ac{SDR} systems. The \ac{NN} complexity could also be further reduced with pruning, and the accuracy could potentially be improved with retraining after quantization and pruning.


\bibliography{refs}
\end{document}